\documentclass[resetfootnote ]{aastex701}

\usepackage{lmodern}
\usepackage{CJKutf8}
\usepackage{amsmath}
\usepackage{hyperref}
\usepackage{xspace}
\usepackage{caption}
\usepackage{upgreek}
\usepackage{ulem}

\newcommand{\suppmattwo}{Appendix\xspace}
\begin{document}
\title{Pinning Down the Geometry of the Type Ic Broad-Line Supernova 2026gzf}

\author[orcid=0009-0007-3401-7133]{Xudong Wen*\begin{CJK*}{UTF8}{gbsn}
(文旭东)\end{CJK*}}
\affiliation{Department of Physics, Tsinghua University, Qinghua Yuan, Beijing 100084, China}
\email[show]{*wenxudong@mail.tsinghua.edu.cn}  

\author[orcid=0000-0002-6535-8500]{Yi Yang*\begin{CJK*}{UTF8}{gbsn}
(杨轶)\end{CJK*}}
\affiliation{Department of Physics, Tsinghua University, Qinghua Yuan, Beijing 100084, China}
\email[show]{*yi\_yang@mail.tsinghua.edu.cn}  

\author[orcid=0000-0002-3900-1452]{Jing Lu\begin{CJK*}{UTF8}{gbsn}
(陆晶)\end{CJK*}}
\affiliation{Department of Physics and Astronomy, Michigan State University, East Lansing, MI 48824, USA}
\email[unshow]{lujing8@msu.edu}

\author[0000-0001-7092-9374]{Lifan Wang}
\affiliation{George P.\ and Cynthia Woods Mitchell Institute for Fundamental Physics \& Astronomy, Texas A\&M University, 4242 TAMU, College Station, TX 77843, USA}
\email[unshow]{lifan@tamu.edu}

\author[0000-0003-1349-6538]{J. Craig Wheeler}
\affiliation{Department of Astronomy, University of Texas, 2515 Speedway, Stop C1400, Austin, TX 78712-1205, USA}
\email[unshow]{wheel@astro.as.utexas.edu}

\author[0000-0003-4663-4300]{Miika Pursiainen}
\affiliation{Department of Physics, University of Warwick, Gibbet Hill Road, Coventry, CV4 7AL, UK}
\email[unshow]{Miika.Pursiainen@warwick.ac.uk }

\author[orcid=0009-0005-2787-9152]{QiuJu Huang}
\affiliation{School of Astronomy and Space Sciences, University of Science and Technology of China, Hefei, 230026, China}
\affiliation{Purple Mountain Observatory, Chinese Academy of Sciences, Nanjing, 210023, China}
\email[unshow]{qjhuang@pmo.ac.cn}

\author[orcid=0000-0003-3635-5375]{Bao Wang}
\affiliation{School of Astronomy and Space Sciences, University of Science and Technology of China, Hefei, 230026, China}
\affiliation{Purple Mountain Observatory, Chinese Academy of Sciences, Nanjing, 210023, China}
\email[unshow]{baowang@pmo.ac.cn}

\author[orcid=0000-0002-6299-1263]{Xuefeng Wu}
\affiliation{Purple Mountain Observatory, Chinese Academy of Sciences, Nanjing, 210023, China}
\affiliation{School of Astronomy and Space Sciences, University of Science and Technology of China, Hefei, 230026, China}
\email[unshow]{xfwu@pmo.ac.cn}

\author[0000-0003-3460-0103]{Alexei~V.~Filippenko}
\affiliation{Department of Astronomy, University of California, Berkeley, CA 94720-3411, USA}
\affiliation{Hagler Institute for Advanced Study, Texas A\&M University, 3572 TAMU, College Station, TX 77843, USA}
\email[unshow]{afilippenko@berkeley.edu}

\author[orcid=0000-0002-0479-7235]{Wolfgang E. Kerzendorf}
\affiliation{Department of Physics and Astronomy, Michigan State University, East Lansing, MI 48824, USA}
\affiliation{Department of Computational Mathematics, Science, and Engineering, Michigan State University, East Lansing, MI 48824, USA}
\email[unshow]{wkerzend@msu.edu}

\author[0000-0002-8597-0756]{Giorgos Leloudas}
\affiliation{DTU Space, Department of Space Research and Space Technology , Technical University of Denmark, Elektrovej 327, 2800 Kgs. Lyngby, Denmark}
\email[unshow]{giorgos@space.dtu.dk}



\author[0000-0001-6797-1889]{Steve Schulze}
\affiliation{Department of Particle Physics and Astrophysics, Weizmann Institute of Science, Rehobot, Israel}
\email[unshow]{steve.schulze@weizmann.ac.il}

\author[0000-0002-0537-3573]{Ferdinando Patat}
\affiliation{European Organisation for Astronomical Research in the Southern Hemisphere (ESO), Karl-Schwarzschild-Str.\ 2, 85748 Garching b.\ M{\"u}nchen, Germany}
\email[unshow]{fpatat@eso.org}

\begin{abstract}
Type Ic broad-line supernovae (SNe Ic-BL) are often associated with energetic explosions that display a prompt outburst of high-energy emission. Since their progenitor lost the H and He envelopes before the explosion exposing the C/O core, their explosion dynamics and geometry can be seen in an unobscured and undistorted way. 
We present imaging polarimetry and spectropolarimetry of the Type Ic-BL SN\,2026gzf obtained 4.6 and 16.5 days after the X-ray shock breakout, which 
was recorded by the Einstein Probe satellite as EP260321a, showing it to be one of the softest and intrinsically dimmest extragalactic fast X-ray transients. 
The persistent low continuum polarization indicates that the outer layer of SN\,2026gzf is mostly spherical, suggesting the explosion did not significantly disrupt the progenitor envelope. 
At day 16.5, the calcium near-infrared triplet displays a peak polarization above 1.5\%. The geometry of the associated line opacity is also compatible with an axisymmetric configuration. The spatial distribution of such oxygen-burning ashes thus indicates the presence of a symmetry axis of the excitation structure within the nearly spherical ejecta. 
The Ca{\sc,II} triplet polarization is dominated by a primary component spanning $\sim25,000\text{--}40,000$,km,s$^{-1}$, while a possible high-velocity polarization substructure of modest statistical significance may indicate a complex, non-axisymmetric excitation geometry in the outer ejecta.
By implementing a three-dimensional Monte-Carlo calculation, we infer that a viewing angle of $\sim 40^{\circ}$ from the symmetry axis of the excitation structure could plausibly reproduce the observed spectral and polarization profiles of the Ca{\sc\,II} triplet.

\end{abstract}

\keywords{\uat{High-energy astrophysics}{739} --- \uat{Stellar astronomy}{1583} --- \uat{Spectropolarimetry}{1973} --- \uat{Type Ic supernovae}{1730}}

\section{Introduction}\label{sec:intro}
The X-ray transient EP260321a was discovered by the Wide-field X-ray Telescope (WXT) onboard the Einstein Probe (EP;~\citealp{2015arXiv150607735Y, 2022hxga.book...86Y}) mission at UTC 2026-03-21 12:30:18 ($T_{\rm trig}$;~\citealp{2026GCN.44068....1H}). Immediate observations with the EP Follow-up X-ray Telescope (FXT) were also performed automatically.
A refined analysis showed that WXT had in fact captured the onset of the transient in the preceding survey observation, with the onset time defined as $T_{0}=$2026-03-21 12:16:08 UTC, or MJD 	61120.511~\citep{2026arXiv260610014Y}. 
Optical follow-up campaigns starting from as early as $\sim T_{0}+1$\,hr detected an evident and rapid brightening of the source (see, e.g., \citealp{2026GCN.44070....1L, 2026GCN.44082....1T, 2026GCN.44089....1S, 2026GCN.44103....1P}). A spectrum obtained at $\sim 12$ hr by the European Southern Observatory Very Large Telescope (ESO VLT) shows a good match to that of Type Ic supernovae with broad absorption lines (SNe~Ic-BL;~\citealp{2026TNSCR1271....1C}).

The early-time evolution of the X-ray luminosity and the spectral energy distribution of EP260321a (SN\,2026gzf) can be well described by thermal-like emission, suggesting a near local thermal equilibrium of the shock. 
Except for the absence of any nonthermal component, such an early evolution is similar to that of SN\,2008D~\citep{2008Sci...321.1185M}, which may be attributed to a nonrelativistic jet, with a shock velocity $v_{\rm shock}\lesssim0.08c$~\citep{2026arXiv260610014Y}. 
The X-ray spectra of WXT and FXT can be well characterized by an absorbed blackbody function with an intrinsic absorption of $N_{\rm H}=8.4_{-1.5}^{+1.6}\times10^{20}$\,cm$^{-2}$), and blackbody temperatures of $kT=124_{-6}^{+7}$\,eV and $112.7_{-2.1}^{+2.1}$\,eV, respectively (see~\citealp{2026arXiv260610014Y} for more details). The soft spectrum within $\sim T_{0}+2200$\,s, together with the rapid decrease of the luminosity at a redshift of $z\approx 0.034$~\citep{2026GCN.44082....1T}, indicates that the early X-ray emission is a plausible candidate for an SN shock breakout. 
No gamma-ray detection was reported for SN\,2026gzf. The temperature and total radiated energy of the event also indicate that it is by far the softest and least luminous event among all extragalactic fast X-ray transients detected by EP~\citep{2025NatAs...9.1073S, 2025arXiv250417034L}, thus strongly disfavoring the existence of an on-axis relativistic jet (see also \citealp{2026arXiv260610014Y, 2026arXiv260610009C, 2026arXiv260610002M, 2026arXiv260609992O, 2026arXiv260610011R}). 

The progenitor star of SN\,2026gzf was inferred to be the core of a massive star with its hydrogen and helium envelopes stripped.
The duration of its shock breakout is significantly longer than the time of a shock breakout from the surface of a Wolf-Rayet (WR) progenitor~\citep{2026arXiv260610009C, 2026arXiv260610014Y, 2026arXiv260610011R, 2026arXiv260610002M}. Modeling of the early X-ray and optical emission of SN\,2026gzf indicates the shock breaking out from an envelope of circumstellar matter (CSM) that extends $\sim 300$ times the solar radii~\citep{2026arXiv260610014Y}. The enrichment of the ambient environment is also compatible with multiple episodes of pre-explosion mass ejections, as indicated by the precursor activity over the past $\sim 12$ yr~\citep{2026arXiv260610009C} before the progenitor's terminal explosion.

Critical information about the explosion geometry is encoded in the polarization spectra. In the first few weeks after the explosion, the rapidly expanding ejecta establishes an electron-scattering atmosphere, with the last scattering of photons reflecting the symmetry of the scattering layers. The continuum level of polarization indicates a deviation from spherical symmetry of the electron-scattering photosphere caused by incomplete cancellation of electric vectors. Polarization measured across various spectral lines traces the distribution of the associated opacity distributions and thus the chemical structures within the SN ejecta~\citep{2008ARA&A..46..433W}. 

Moreover, polarimetry of stripped-envelope SNe offers the most direct insight into the explosion mechanism because it is not hidden below a massive H/He envelope. The latter has been blown away by a stellar wind or stripped off by a close companion star (see, e.g., ~\citealp{1994Natur.371..227N, 1995ApJ...448..315W, 2010ApJ...725..940Y, 2012Sci...337..444S, 2024NatCo..15.7667S}). Therefore, the shape of the early ejecta preserves the information of the initial shock breakout and hence provides geometric diagnostics of the explosion model. 
The formation mechanism of polarization in SNe~Ic and Ic-BL  is still unclear. Mass transfer to a close companion~\citep{2014AJ....148...68B} and nonspherical explosions~\citep{2013ApJ...779...60M} could be possible explanations. 
Polarimetric probes of SNe~Ic have been carried out only in relatively few cases, including the GRB-associated SNe\,1998bw~\citep{2001ApJ...555..900P}, 2003dh~\citep{2003Natur.426..157G}, and 2006aj~\citep{2006A&A...459L..33G, 2011ASPC..449..421G}, as well as events without a GRB detection such as SNe\,2002ap~\citep{2002PASP..114.1333L, 2002ApJ...580L..39K, 2003ApJ...592..457W}, 2005bf~\citep{2007MNRAS.381..201M}, 2007gr~\citep{2008ApJ...689.1191T}, and 2014ad~\citep{2017MNRAS.469.1897S}. 
Detailed investigations through numerical simulations also suggest that nonaxisymmetric three-dimensional (3D) geometry is ubiquitous in stripped-envelope SNe~\citep{2012ApJ...754...63T, Tanaka2017}.

Here we report spectropolarimetry of the Type Ic-BL SN\,2026gzf at 16.5 days after the initial X-ray shock breakout\footnote{Throughout the paper, all phases are given relative to the X-ray shock breakout at UTC 2026-03-21 12:16 or MJD 61120.511~\citep{2026arXiv260610014Y}.}.
In Section~\ref{sec:specpol} we present the observations and the identification of a symmetry axis of the Ca{\sc\,II} near-infrared triplet (hereafter Ca{\sc\,II}\,NIR3) opacity. Physical properties of the SN ejecta are also characterized using the radiative-transfer code {\sc TARDIS}~\citep{2014MNRAS.440..387K}. Section~\ref{sec:model} models the Ca{\sc\,II}\,NIR3 polarization with a 3D Monte-Carlo calculation and interprets the spatial distribution of the line opacity.   Discussion and concluding remarks are given in Section~\ref{sec:discussion}.

\section{Imaging and Spectropolarimetry of SN~2026gzf}~\label{sec:specpol}
\textcolor{black}{Imaging polarimetry of SN\,2026gzf was obtained with the IPOL mode of the FOcal Reducer and low-dispersion Spectrograph 2 (FORS2;~\citealp{1998Msngr..94....1A}) on UT1 (Antu) of the ESO VLT on UTC 2026-03-26 02:36 / MJD\,61125.109 (Prog.\,ID 116.28T2.002, PI M. Pursiainen).}
We also requested Director’s Discretionary Time observations with FORS2 to conduct a spectropolarimetric probe of the ejecta geometry of SN\,2026gzf (Prog.\,ID 116.2ASW.001, PI Y. Yang). The spectropolarimetry data presented in this paper were obtained on UTC 2026-04-07 01:00 / MJD\,61137.042. Observations were carried out in the Polarimetric Multi-Object Spectroscopy (PMOS) mode, with the 300V grism in place and coupled to a 1$''$-wide slit. 
All integrations were done when the SN was close to the meridian and at low airmass levels ($\lesssim$1.1), allowing for effective compensation of blue light loss using the linear atmospheric dispersion compensator \citep{1982PASP...94..715F, 1997SPIE.2871.1135A}). 
The configuration provides a wavelength coverage of $\sim 3460$--9300\,\AA\ and a spectral resolution near the central wavelength (5849\,\AA) of $R\approx440$, corresponding to a width of $\sim 13.3$\,\AA\ per  resolution element. Considering the negligible effect on the polarization signal due to contamination by second-order light (see the Appendix of~\citealp{2010A&A...510A.108P}), we did not use the GG435 order-sorting filter, which has a cutoff wavelength of $\sim 4350$\,\AA.

Details of the FORS2 IPOL and PMOS data reduction and the derivation of the Stokes parameters can be found in the FORS2 Spectropolarimetry Cookbook and Reflex Tutorial\footnote{\url{https://ftp.eso.org/pub/dfs/pipelines/instruments/fors/fors-pmos-reflex-tutorial-1.9.pdf}}, \citet{2017MNRAS.464.4146C}, \citet{2023A&A...674A..81P}, and Appendix~A of~\citet{2020ApJ...902...46Y}, 
\textcolor{black}{
following the procedures described by~\citet{2015ApJ...815L..10L}, \citet{2017ApJ...837L..14L}, \citet{2006PASP..118..146P}, and \citet{2007MNRAS.381..201M}. Because the sensitivity of VLT FORS2 decreases rapidly below $\sim 4000$\,\AA\ and leads to a dramatic increase in polarization uncertainty at the blue end of the optical spectrum, we made no attempt to characterize the polarization features below $\sim 4000$\,\AA. 
A log of the VLT observations is provided in Table.~\ref{Table_pol}.}
The flux spectra were extracted using the standard extraction algorithm following the FORS2 reduction procedure.

\begin{figure*}[ht]
    \centering
    \includegraphics[trim={0.0cm 0.0cm 0.0cm 0.0cm},clip,width=1.0\textwidth]                                {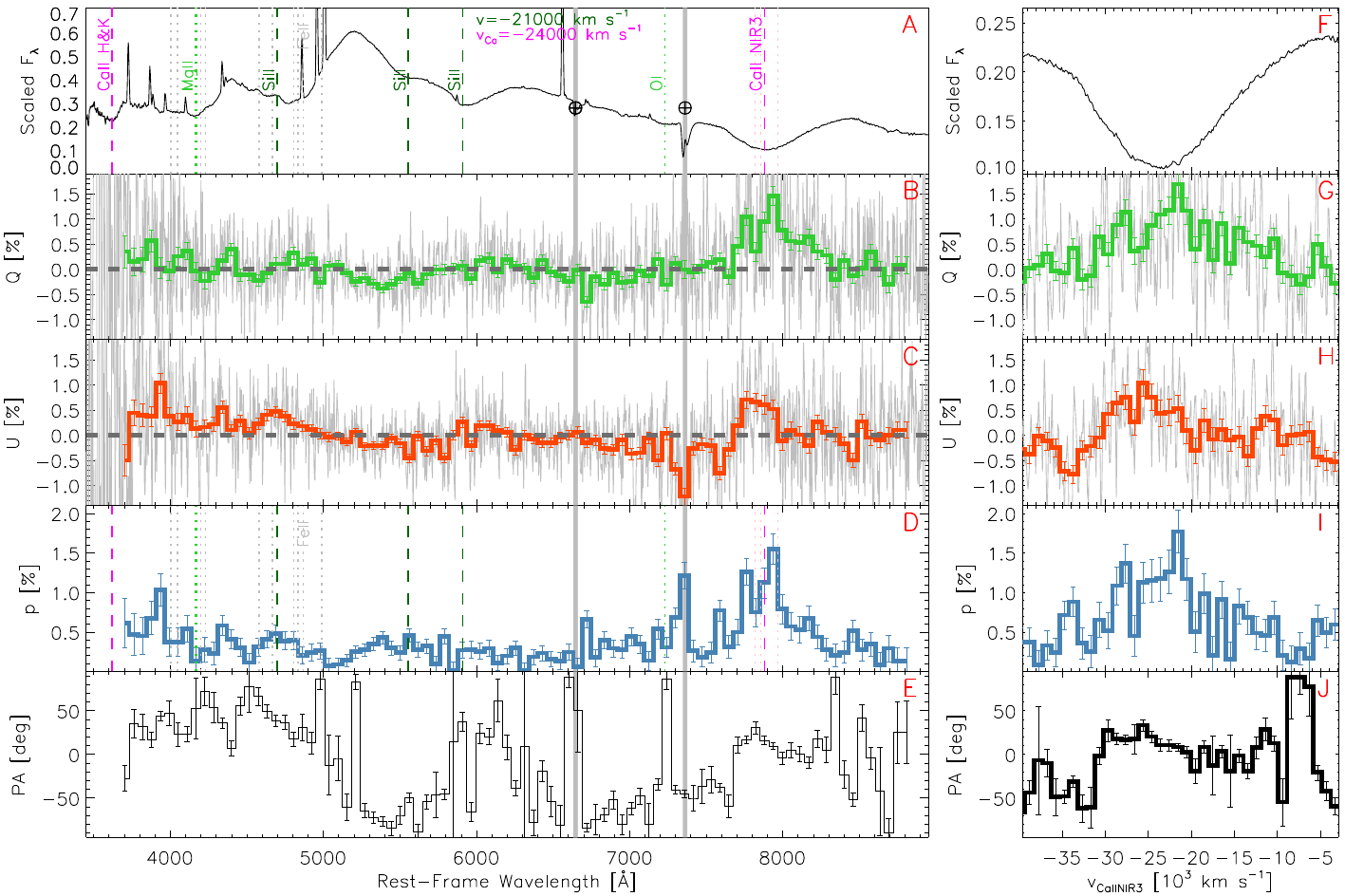}
    \vspace{-0.0 cm}
    \caption{Spectropolarimetry of SN\,2026gzf at 16.5 days after the X-ray shock breakout. The {\it left column} (from top to bottom) shows (A) the arbitrarily scaled total-flux spectrum with major spectral lines identified (telluric features are indicated with an Earth symbol; all narrow spikes originate from an underlying nebulosity); (B, C) the intensity-normalized Stokes parameters $\rm Q$ and $\rm U$, with zero level indicated by the horizontal gray dashed lines; (D) the polarization spectrum ($\rm p$); and (E) the polarization position angle. 
    A 60\,\AA\ bin size is used in Panels B--E for clarity. 
    The {\it right column} displays the Ca{\sc\,II}\,NIR3 absorption profile; panels G--J show the polarization as a function of blueshifted velocity relative to the central rest wavelength of  Ca{\sc\,II}\,NIR3 at 8567\,\AA\ using 30\,\AA\ bins for clarity. 
    The light gray curves show the Stokes $\rm Q$ and $\rm U$ parameters sampled under the VLT FORS2 pixel scale resulted from our instrument setup, namely $\approx$3.3\,\AA\,pixel$^{-1}$.
    The $\rm p$ displayed in (D, I) are the debiased polarization degree, calculated using Equation~\ref{eq:debiased_polarization}.
    }
~\label{fig:iqu_ep1}
\end{figure*}

Figure~\ref{fig:iqu_ep1} presents the observed flux and polarization spectra of SN\,2026gzf on day 16.5, covering a rest-frame wavelength range of 3700--8800\,\AA. The data displayed in the left and the right panels have been rebinned to 60\,\AA\ and 30\,\AA, respectively, with the latter capturing the polarization profile of Ca{\sc\,II}\,NIR3 in order to increase the signal-to-noise ratio (S/N) but also to ensure that major line structures are adequately sampled. All narrow spikes in the flux spectrum (Figure~\ref{fig:iqu_ep1}A) originated from the underlying host, which displays no identifiable offset as seen from the 2D raw spectropolarimetry data.

\subsection{Interstellar Polarization}~\label{sec:specpol_isp}

Dust grains along the SN-Earth line of sight (LOS) induce interstellar polarization (ISP), manifested as a wavelength-dependent baseline shift of the Stokes $\rm Q$ and $\rm U$ parameters, which can be empirically characterized by Serkowski's law~\citep{1975ApJ...196..261S}. An upper limit on the Galactic ISP component yields $p_{\rm ISP}^{\rm MW} [\%] < 9 \% \times E(B-V)$~\citep{1975ApJ...196..261S}. The 3D extinction map estimates of the Galactic reddening toward SN\,2026gzf imply $E(B-V) = 0.021$\,mag~\citep{1989ApJ...345..245C, 2011ApJ...737..103S}, indicating $p_{\rm ISP}^{\rm MW} \lesssim0.19$\%. 
\Citet{2026arXiv260610011R} placed an upper limit of the host extinction of $A_{V}\lesssim0.13$\,mag based on its correlation with the equivalent width of the Na{\sc\,ID} absorption line~\citep{1997A&A...318..269M, 2012MNRAS.426.1465P, 2013ApJ...779...38P, 2018A&A...609A.135S}. This corresponds to a $p_{\rm ISP}^{\rm Host} \lesssim0.38$\% assuming the LOS dust within the SN host follows a Galactic $R_{V}=3.1$ extinction law~\citep{1989ApJ...345..245C}. Moreover, the absence of any wavelength dependence in the observed polarization spectrum of SN\,2026gzf suggests no strong ISP along the SN-Earth LOS.
We therefore assume negligible ISP in further analysis.

\subsection{Continuum Polarization}~\label{sec:specpol_cont}

After correcting the observed $\rm Q$ and $\rm U$ spectra from a low-level wavelength-dependent instrumental polarization ($\lesssim$0.1\%~\citep{2007ASPC..364..503F, 2014A&A...561A..82S, 2017MNRAS.464.4146C}, the observed polarization degree $\rm{p_{\rm obs}}$ and position angle (PA$_{\rm obs}$) can then be expressed as: 
\begin{equation}
    \rm{p_{\rm obs}}=\sqrt{Q^{2}+U^{2}}, \  {\rm PA_{obs}}=\frac{1}{2}\tan^{-1} \bigg{(} \frac{U}{Q} \bigg{)},
\end{equation}
Since $\rm{p_{\rm obs}}$ is not negative by definition, errors in $\rm Q$ and $\rm U$ thus always lead to higher values compared to the true degree of polarization $\rm p$. 
We therefore debias the observed polarization spectra by calculating 
\begin{equation}
\begin{aligned}
\rm{p_{\rm }} = (p_{\rm obs} - \sigma_{p}^2 / p_{\rm obs}) \times h(p_{\rm obs} - \sigma_{p}), \ 
{\rm PA} = {\rm PA}_{\rm obs}.
\end{aligned}
\label{eq:debiased_polarization}
\end{equation}
Unless otherwise stated, all polarization degrees plotted and analyzed throughout this work refer to the debiased polarization $\rm p$, rather than $\rm p_{obs}$.

Broad-band polarization of SN\,2026gzf on day 4.6 is consistent with zero (see Table~\ref{Table_pol}).
We also estimate the continuum polarization on day 16.5 by calculating the error-weighted mean polarization over 6400--7100\,\AA. Without any obvious spectral features, our measurement suggests 
$\rm Q_{\rm Cont}^{\rm +16.5\,d} = -0.16\pm0.21$\,\%, 
$\rm U_{\rm Cont}^{\rm +16.5\,d} = -0.15\pm0.13$\,\%, which is similar to the near-peak continuum polarization of the Type Ib/c SN\,2008D~\citep{2009ApJ...705.1139M} measured across the range  7100--7500\,\AA. 
The observed polarization of the continuous spectrum is $\rm p_{\rm obs}\sim0.22\%$, and after bias correction, the continuous spectrum is nearly unpolarized ($\rm p\sim0.08\%$).
For a radial density distribution following a power law, $n_{r}\propto r^{-6}$, a nearly spherical configuration, although a slight intrinsic asphericity cannot be excluded for viewing angles close to the symmetry axis (see Appendix~\ref{sec:Continum Polarization}).

\setlength{\tabcolsep}{5pt}
\begin{table}[ht]
\begin{center}
\caption{VLT Polarimetry of SN\,2026gzf.~\label{Table_pol}}
\begin{normalsize}
\begin{tabular}{c|c|c|c|c|c}
\hline
\hline
Mode  &  MJD-OBS / Phase$^{a}$  &  Exp. Time$^{b}$  & Filter / Spectral Range$^{c}$ &  $\rm Q^{\rm Cont}$ & $\rm U^{\rm Cont}$  \\
 \#   &  [day]      &  [s]       &  ($\lambda_{0}$, FWHM)   &  [\%]           &  [\%]     \\
\hline
IPOL  & 61125.109 / day 4.6  &  $1\times4\times60$          & b$_{\rm HIGH}$ (4370\,\AA, 1020\,\AA) &  $0.23\pm0.15$  &  $0.05\pm0.14$ \\
      & 61125.114    &  $1\times4\times45$          & v$_{\rm HIGH}$ (5550\,\AA, 1232\,\AA) &  $-0.21\pm0.12$  &  $0.12\pm0.13$ \\
      & 61125.119   &  $1\times4\times45$          & R$_{\rm SPECIAL}$ (6550\,\AA, 1250\,\AA) &  $0.04\pm0.12$  &  $0.13\pm0.12$ \\
      & 61125.123    &  $1\times4\times45$          & I$_{\rm BESS}$ (7680\,\AA, 1380\,\AA) &  $0.02\pm0.18$  &  $0.33\pm0.18$ \\
\hline
PMOS  & 61137.042 / day 16.5  &  $2\times4\times480$  &  6400--7100\,\AA\ &  $-0.16\pm0.21$ & $-0.15\pm0.13$   \\
\hline
\hline
\end{tabular}\\
{$^{a}$Relative to the onset time of EP260321a on UTC 2026-03-21 12:16:08 / MJD 61120.511. } \\
{$^{b}$Each set of observations consists of $n$ [loops]$\times$4 [half-wave plate angles]$\times$[time of integration].} \\
{$^{c}$Denoted as the central wavelength $\lambda_{0}$ and the full width at half-maximum intensity filter transmission.}
\end{normalsize}
\end{center}
\end{table}

\subsection{An Overall Axisymmetric Excitation Structure}~\label{sec:specpol_line}
\begin{figure*}[ht]
     \includegraphics[width=1.0\linewidth]{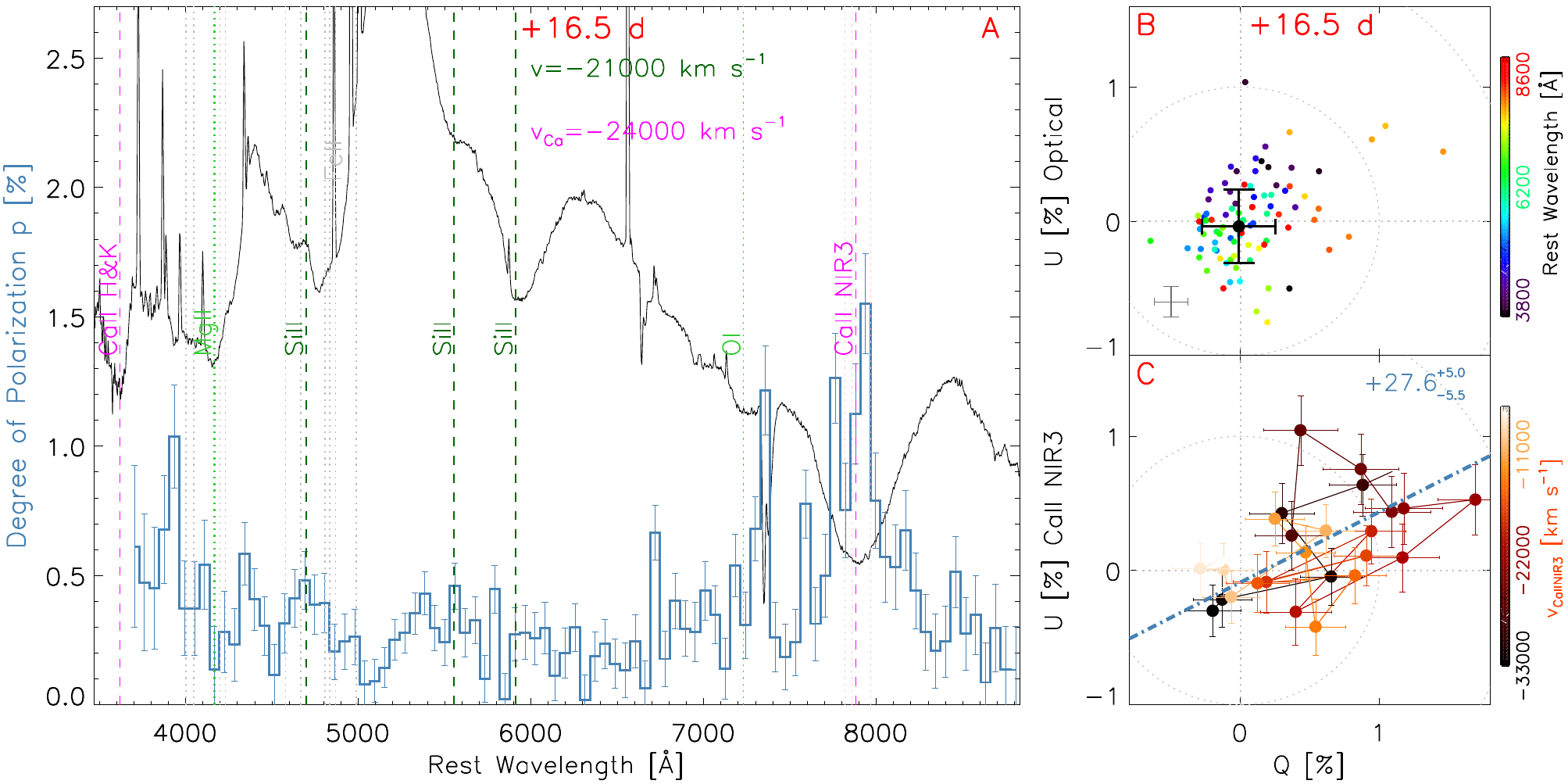}
     \caption{{\it Panel A:} intrinsic polarization of SN\,2026gzf at day 16.5.  
     The debiased degree of polarization, $\rm p$, calculated using Equation~\ref{eq:debiased_polarization}, is shown as a navy histogram with 60\,\AA\ binning, together with the full-resolution, arbitrarily scaled flux spectrum ($F_{\lambda}$; black line).
     All narrow spikes originate from an underlying nebulosity. Vertical lines identify several major spectral features. 
     {\it Panel B:} polarization of SN\,2026gzf displayed on the Stokes $\rm Q-U$ plane. The top panel presents the data between rest-frame wavelengths of 3800 and 8800\,\AA, with a 60\,\AA\ bin size adopted for clarity. The wavelength of each bin can be read from the color bar at the right. The solid black circle close to the origin identifies the continuum polarization deduced over the wavelength range of 6400--7100\,\AA. Panel C has a similar layout as the top panel but is restricted to the Ca{\sc\,II}\,NIR3 profile from blueshifted velocities ranging from 33,000 to 3000 km\,s$^{-1}$ with a 30\,\AA\ spectral binning. The navy dotted-dashed line is a fit to the displayed data points, whose orientation in degrees with uncertainties is indicated by the navy text.
     }~\label{fig:pol}
\end{figure*}

At day 16.5, the Ca{\sc\,II}\,NIR3 feature of SN\,2026gzf exhibits prominent polarization that peaks at $\sim 1.6$\% (Figures~\ref{fig:iqu_ep1} and~\ref{fig:pol}A). We also present the observed polarization on the Stokes $\rm Q-U$ plane in Figure~\ref{fig:pol}B, which offers an intuitive inspection of the Stokes parameters across the continuum and various spectral features~\citep{2001ApJ...550.1030W}. At a given wavelength, the polarization degree and position angle are given by its distance to the origin and slope, respectively. 
The varying distances of the data points from the origin reflect wavelength-dependent polarization arising from different line and continuum formation regions, which probe different optical depths and geometric structures within the ejecta.
A perfect axially symmetric structure maintains a universal polarization position angle, thus manifesting a straight line on the $\rm Q-U$ plane that is known as the dominant axis~\citep{2003ApJ...591.1110W, 2010ApJ...722.1162M},  
\begin{equation}
    \rm U =\alpha + \beta Q. 
\end{equation}
Deviations from such a ``dominant axis'' imply the breaking of axisymmetry, indicative of the presence of clumps or other substructures. 

As shown in Figure~\ref{fig:pol}B, except for the blueshifted Ca{\sc\,II}\,NIR3 feature that spans rest wavelenth 7500--8500\,\AA, polarization across the entire observed wavelength range is clustered around the origin, displaying no preferred orientation on the $\rm Q-U$ plane. 
On the contrary, Polarization over the Ca{\sc\,II}\,NIR3 feature is overall consistent with a dominant axis obtained by performing an error-weighted linear least-squares ﬁt to the data points across the line (Figure~\ref{fig:pol}C), suggesting an overall axisymmetric Ca{\sc\,II} opacity distribution above the electron-scattering photosphere. 

We also note that only Ca{\sc\,II}\,NIR3 instead of other features of intermediate-mass elements (IMEs) displays such a signature. The low polarization across other spectral features, such as Si{\sc\,II}\,$\lambda$6355, O{\sc\,I}\,$\lambda$7774, and Mg{\sc\,II}\,$\lambda$4481, can be understood as a rather  
uniform excitation structure to produce these lines at the location of the electron-scattering photosphere on day 16.5. 
The temporal evolution of the O{\sc\,I}\,$\lambda7774$ and Ca{\sc\,II}\,NIR3 features in SNe~Ic encodes the ionization structure of the ejecta in a physically unified way. The ionization potentials of neutral oxygen and neutral hydrogen are nearly identical, respectively 13.62\,eV and 13.60\,eV. 
This implies that in the H-free ejecta of an SN~Ic, oxygen recombination plays a role analogous to hydrogen recombination in Type II SNe by regulating the electron-scattering opacity near the photosphere.
Below this front, oxygen is predominantly ionized, contributing free electrons that maintain a high electron-scattering opacity. Above and at the front, neutral oxygen is present, giving rise to the observed O{\sc\,I} absorption in a narrow zone immediately above the photosphere.
As the SN brightens toward maximum light, the rising ultraviolet flux drives rapid photoionization of this neutral oxygen layer, causing the O{\sc\,I} feature to weaken or disappear. 
Consequently, the O{\sc\,I}\,$\lambda7774$ is commonly used as an approximate tracer of the photospheric velocity in Type Ic SNe, although its exact formation depth depends on the line optical depth and the detailed ionization structure of the ejecta~\citep{Dessart2015,Dessart2016,Fremling2018,Liu2016}. 
The absorption minimum of O{\sc\,I}\,$\lambda7774$
corresponds to an velocity of approximately
21,000\,km\,s$^{-1}$, close to the inner-boundary
velocity of $v_{\rm inner}=18,807^{+110}_{-120}$\,km\,s$^{-1}$
in the TARDIS model at the same epoch
(Appendix~\ref{sec:tardis}). It should be clarified that the TARDIS values are constrained by the global spectral fit, which does not fully reproduce the detailed shape of the O{\sc\,I}\,$\lambda7774$ spectral line.

Because Ca{\sc\,II} requires a rather low excitation energy ($\chi=1.7$\,eV), compared to that of Si{\sc\,II} ($\chi=8.12$\,eV), and remains strong at lower densities compared to that of Si{\sc\,II}, it provides a sensitive tracer of the excitation front as sculpted by energy deposition.  
The overall axisymmetric excitation structure as traced by the dominant axis of the Ca{\sc\,II}\,NIR3 polarization can be induced by an axisymmetric energy deposition by the distribution of nickel that heats the ejecta from within the inner layers. 
Therefore, the polarization at day 16.5 can be naturally attributed to an elongated pattern of energy deposition. 
This inferred structure could be compatible with a bipolar, jet-induced explosion, which does not involve a dramatic disruption of the stellar envelope of the progenitor.

\subsection{Potential Substructures of the Ca{~II}~NIR3 Opacity Distribution}~\label{sec:specpol_substructure}
A more careful inspection of the Ca{\sc\,II}\,NIR3 polarization is shown in Figure~\ref{fig:iqu_ep1}, Panels F--J. With a 30\,\AA\ binning, two peaks in the polarization spectra can be identified at blueshifted velocities of $\sim 28,000$ and 22,000\,km\,s$^{-1}$. When displayed on the $\rm Q-U$ plane, although this line polarization can be fitted by a straight line, a departure of the modulation perpendicular to the dominant axis can also be seen at a velocity of $\sim 28,000$\,km\,s$^{-1}$ (see Figure~\ref{fig:pol}C). 
We remark that although our inspection of the Ca{\sc,II},NIR3 polarization profile indicates a $\gtrsim3\sigma$ departure from the dominant axis at 28,000\,km\,s$^{-1}$, the corresponding deviation from the local polarization trend defined by the neighboring velocity bins is approximately $1.8\sigma$. Given the limited S/N, the identification of any substructure across the line should therefore be regarded as tentative.
Accordingly, any interpretation of this possible secondary component in terms of a complex excitation or opacity structure remains speculative. Irrespective of whether this local variation is physical, the overall approximately axisymmetric configuration inferred from the full Ca II NIR3 polarization profile remains unchanged. Future spectropolarimetric observations at higher spectral resolution and S/N will be essential for resolving potential substructures across the line profile and placing more stringent constraints on the associated ejecta geometry.

\begin{figure*}[htbp]  
    \centering
    \includegraphics[trim={0.0cm 0.0cm 0.0cm 0.0cm},clip,width=1.0\textwidth]{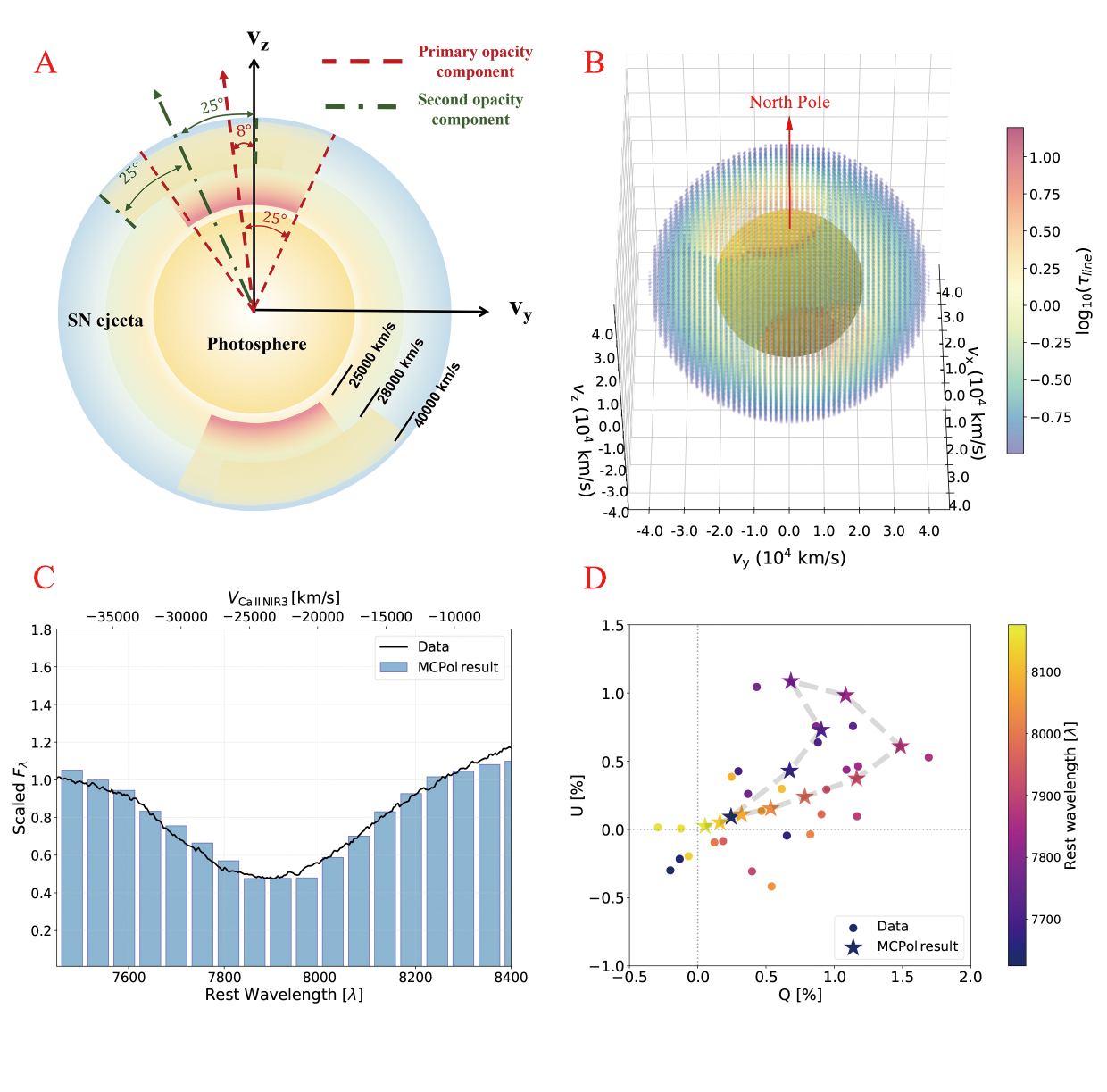}
    \vspace{-1.0cm}
    \caption{
    {\it Panel A:} Schematic illustration of the two-component Ca{\sc\,II} opacity distribution. Symmetry axes of the primary (25,000--40,000~km~s$^{-1}$) and secondary (28,000--40,000~km~s$^{-1}$) components, which are located above the photosphere, are indicated by red and green arrows, respectively. The symmetry axis of the primary component has a $8^{\circ}$ inclination angle relative to the $v_{z}$ axis,  
    while the secondary component has an inclination angle relative to the $v_{z}$ axis of $25^{\circ}$. The difference between the two components is thus 17$^{\circ}$. 
    {\it Panel B:} 3D map of the Ca{\sc\,II} opacity viewed at a $40^\circ$ angle relative to the $v_{z}$-axis. The radius of the inner sphere centered at the origin indicates the size of the photosphere. The optical depth of the Ca{\sc\,II} is presented in a logarithmic scale and color-coded as indicated by the color bar on the right-hand side. Overlapping primary and secondary components create a high-optical-depth region covering the northern pole, where the opacity increases radially inward. Southern components are obscured behind the photosphere, visible as the dark-red shaded area in the lower hemisphere.
    {\it Panel C:} The model-synthesized flux spectrum of Ca{\sc\,II}\,NIR3 (blue histograms) for the opacity distribution illustrated in Panel B compared to the observation (black curve). Both spectra have been normalized to match the continuum flux measured from adjacent wavelength ranges.
    {\it Panel D:} The model polarization modulation across the Ca{\sc\,II}\,NIR3 line profile induced by the opacity distribution presented in Panel B compared to the observations shown in Figure~\ref{fig:pol}C. Polarization at different wavelengths is indicated by the color bars to the right.}~\label{fig:model} 
\end{figure*}

\section{Three-dimensional Monte-Carlo Modeling of the Ca~II Opacity Distribution}~\label{sec:model}
We utilize the 3D Monte-Carlo Polarization Simulation Code (MCPol;~\citealp{2023ApJ...955....9W}) to model the Ca{\sc\,II}\,NIR3 polarization on day 16.5. 
The code tracks photon packets undergoing electron and line-resonance scattering in the SN ejecta to generate flux and polarization spectra for different observer LOS.
The low continuum polarization implies that a broad range of slight aspherical configurations become nearly degenerate with a spherical photosphere when projected onto the plane of the sky. We therefore adopt a spherical photosphere as an effective approximation to minimize the parameter degeneracy associated with the unknown viewing angle and intrinsic axial ratio. We follow \citet{Tanaka2017} and construct an idealized model in which the line opacity is locally enhanced above a spherical photosphere. This configuration naturally reproduces the observed polarization features.
While the model is not unique and alternative parameter combinations may produce similar results, it is intended to identify the simplest effective line-opacity distribution capable of reproducing the observed spectropolarimetric signatures, rather than to uniquely reconstruct the intrinsic three-dimensional Ca II geometry.

The spherically symmetric electron-scatting atmosphere on day 16.5 justifies our assumption of homologously expanding SN ejecta. Unpolarized photon packets are emitted at the inner boundary of the ejecta ($r_{\rm in}$), which is located at an electron-scattering optical depth of $\tau_{\rm in} = 3$, consistent with previous studies~\citep{1991A&A...246..481H,Kasen2003, Hole2010}. The radius of the photosphere $r_{\rm ph}$ is defined to be where the electron-scattering optical depth reaches unity ($\tau_{\rm es} = 1$). 
The density is assumed to follow a power-law profile ($\rho\propto r^{-n}$) with an index of $n=6$, consistent with typical values ($n\approx 6$--7) adopted for the line-forming regions of stripped-envelope SNe (see, e.g.,~\citealp{2000ApJ...534..660I, 2000ApJ...545..407M}, and the {\sc TARDIS} spectral-fitting results (Appendix~\ref{sec:tardis}). 

For the line-scattering region above the photosphere, we adopt the Sobolev approximation to treat line interactions~\citep{1970MNRAS.149..111C}, which is well justified in SN ejecta with large velocity gradients~\citep{Jeffery1989}. 
To account for the overall axisymmetric opacity distribution of Ca{\sc\,II}\,NIR3 including a potentially high-velocity substructure (see  Section.~\ref{sec:specpol_substructure} and Figure~\ref{fig:pol}), we construct a 3D, two-component geometric model in velocity space. 
This structure comprises primary and secondary line-opacity components, the velocities of which span 25,000--40,000~km~s$^{-1}$ and 28,000--40,000~km~s$^{-1}$, respectively. 
In the appendix~\ref{sec:angle}, we also discuss a case featuring only a dominant opacity component distribution, without secondary structural components.
We parameterized the Ca{\sc\,II}\,NIR3 profiles by treating the $\lambda8542$ optical depth ($\tau_{\rm line}$) as a free parameter, while the $\lambda8662$ and $\lambda8498$ opacities are scaled by factors of $1/1.8$ and $1/10$ ~\citep{Kasen2003}. 

The spatial distribution of the line opacities of both components can be described as
\begin{equation}
    \tau_{\rm line}(r)=\tau_{0} 
    \bigg{(}\frac{r/t}{25,000\,\mathrm{km\,s^{-1}}} \bigg{)}^{-n_{\mathrm{line}}}F(\theta), 
\end{equation}
where $\tau_{0}$ denotes the benchmark optical depth of the line opacity measured at the radius of the photosphere at day 16.5. The radial distance from a given point to the center of the photosphere is represented by $r$, and $t$ gives the time after the SN explosion. 
We parameterize the angular dependence of the line opacity by introducing an asymmetry factor $F(\theta)$, which enhances the line opacity by a factor of $f_{\tau}$ within a region with a half-angle of $\theta_{0}$, 
\begin{equation}
F(\theta) = 1 + (f_{\tau} - 1) \Theta(\theta_0 - \theta),
\end{equation}
where $\Theta$ represents the Heaviside step function. 
In this work, we calculate the Ca{\sc\,II}\,NIR3 polarization assuming an enhancement by a factor of $f_{\tau}=10$ for both the primary and secondary opacity components, each manifesting a cone-like configuration, with an opening half angle of $\theta = 25^{\circ}$. To account for the departure of the secondary  component from the dominant axis (Figure~\ref{fig:pol}), the symmetry axes of these two components are set to be misaligned, with inclination angles of $\phi_{1} = 8^{\circ}$ and $\phi_{2} = 25^{\circ}$ measured with respect to the $z$ axis, respectively (see Figure~\ref{fig:model}a).  
Along the radial direction, the Ca{\sc\,II} opacity within the enhanced cone also scales with the electron density at any given position of the SN ejecta, which follows an $n=6$ power-law distribution. A benchmark optical depth of $\tau_{\mathrm{0}} \approx 1.6$ was set for both components.

For a line-opacity distribution deviating from spherical symmetry, both the shape of its spectral line and the polarization modulation across the feature, which can be seen as the morphology of the trajectory on the $\rm Q-U$ plane, will be strongly dependent on the viewing angle (see Appendix~\ref{sec:angle}). 
By simultaneously fitting the flux and polarization modulations across the Ca{\sc\,II}\,NIR3 line profile, we infer a $\sim40^{\circ}$ inclination angle of the symmetry axis of the primary bipolar opacity enhancement measured with respect to the LOS. 
Schematic drawings of the two-component opacity distribution and the 3D distribution of the Ca{\sc\,II} are presented in Figures~\ref{fig:model}A and \ref{fig:model}B, respectively. The associated model-synthesized line profile and polarization across the spectral feature are respectively in Figures~\ref{fig:model}C and  \ref{fig:model}D.

We note that the opacity enhancement near the south pole that falls on the far side of the SN ejecta will be blocked by the electron-scattering atmosphere (Figure~\ref{fig:model}a). Therefore, the opacity distribution toward the far side of the SN ejecta remains unprobed. 
The geometric configuration implemented in our model is equivalent to an extended region of significantly enhanced Ca{\sc\,II} line opacity localized near the north pole, rather than a bipolar symmetric enhancement. Nevertheless, the conclusion regarding the macroscopic enhancement of Ca{\sc\,II} line opacity near the pole region remains robust, as a uniform distribution of numerous small clumps fails to effectively break the geometric symmetry, making it highly incapable of producing the observed strong line-polarization features~\citep{Tanaka2017}. 
It should be emphasized that our conclusions do not depend on the specific parameterization adopted in this work. Rather, any model that reproduces the same large-scale distribution of the effective Ca II line opacity along the LOS is expected to produce similar polarization signatures and should therefore be considered an equally viable explanation of the current observations\citep{Jeffery1989, Dessart2011}.

\section{Discussion}~\label{sec:discussion}

\subsection{A Spatially Elongated Excitation Structure Revealed by Ca~II Polarization}
The strong polarization that manifests an overall well-defined dominant axis across the Ca{\sc\,II}\,NIR3 clearly indicates an axisymmetric excitation structure within the ejecta of SN\,2026gzf. The low polarization across the continuum and other prominent spectral lines suggests that the axisymmetric structure revealed by the Ca{\sc II},NIR3 polarization is restricted to the Ca{\sc II} line-forming region and does not extend into either the electron-scattering photosphere or the line-forming regions of the other major species.
The absence of Ca{\sc\,II}\,NIR3 in the first flux spectrum of SN\,2026gzf at day 2.4 and its strengthening with time (Martin-Carrillo et al. 2026; Rastinejad et al. 2026) is a natural counterpart of the receding excitation front within the expanding and cooling ejecta. 
Calcium at the highest observed velocities in the ejecta is most plausibly identified as \textit{primordial, unburned calcium}
present in the outer layers of the progenitor prior to the explosion, rather than as nucleosynthetic calcium produced by oxygen burning, which is confined to the deeper, slower-moving ejecta. Within the first few days after the SN explosion, this primordial calcium exists predominantly as Ca$^{2+}$ due to the hard radiation field, rendering the Ca{\sc\,II} features weak or absent. 
As the ejecta expand and the radiation field softens, the recombination Ca$^{2+} \rightarrow$ Ca$^+$ proceeds, and the Ca{\sc\,II} features strengthen progressively. 

Polarization of the Ca{\sc\,II}\,NIR3 thus reflects the geometry of the Ca$^+$ recombination zone rather than purely the calcium abundance distribution. 
An asymmetric explosion or asymmetric energy deposition by radioactive $^{56}$Ni can produce an asymmetric radiation field, which in turn drives an asymmetric ionization front. 
The subsequent spectral evolution yields a strong Ca{\sc\,II}\,NIR3 feature at later times, thus reflecting the combined evolution of two ionization fronts rather than the simple recession of the photosphere through nucleosynthetic layers between different compositions. 
The two components of the Ca{\sc\,II}\,NIR3 polarization of SN\,2026gzf may therefore probe distinct geometric structures: the high-velocity component from primordial calcium in the outer progenitor envelope traces its geometry and interaction with the propagating shock, while the lower-velocity component traces the geometry of the inner explosive nucleosynthesis. 
The misalignment between these two components, as suggested by departures from a single dominant axis on the Stokes $\rm Q-U$ plane, may then reflect the angle between the elemental structure within the progenitor envelope and the excitation front induced by the SN explosion, 
rather than the presence of localized ionization bubbles or clumps within the ejecta. 
A comprehensive physical model of Type Ic SN spectropolarimetry must therefore account for the ionization-front geometry as a primary contributor to both the spectral and polarimetric signatures, instead of treating observed line opacity distributions as direct tracers of elemental abundance geometry.

\subsection{Viewing Angle Inferred From the Polarized Ca~II and the Weak O~I Lines}
The weak O{\sc\,I}\,$\lambda$7774 line of SN\,2026gzf deserves further attention. 
We note that the spectropolarimetry of the Type Ic-BL SN\,2014ad, which also shows no GRB counterpart, has identified loop-like patterns in both O{\sc\,I}\,$\lambda$7774 and Ca{\sc\,II}\,NIR3 opacities around its peak luminosity~\citep{2017MNRAS.469.1897S}. 
On the contrary, the O{\sc\,I}\,$\lambda$7774 absorption line of SN\,2026gzf on day 16.5 exhibits a considerably shallower spectral profile (see also, e.g.,~\citealp{2026arXiv260609992O, 2026arXiv260610011R}). 
Such a difference can naturally be attributed to the deviation from spherical symmetry in the excitation front within the SN ejecta, which traces the energy propagation and thus the explosion geometry of the progenitor star. In particular, O{\sc\,I} and Ca{\sc\,II} features yield major indicators of unburned oxygen and oxygen-burning ashes in the outer ejecta layers, respectively. 
The polarization and spectral profiles of these two lines would be sensitive to the viewing angle. 

The time-variant strong O{\sc\,I} at early phases followed by strong Ca{\sc\,II} at later phases was first clearly identified and documented in the spectropolarimetric observations of SN\,2002ap~\citep{2003ApJ...592..457W}, who detected a prominent polarized O{\sc\,I}$\,\lambda$7774 feature moving at $\sim$20,000\,km\,s$^{-1}$ at pre-maximum-light epochs. 
In contrast, SN\,2026gzf exhibits a considerably shallower and unpolarized O{\sc\,I} near its peak brightness. One possible scenario to account for such  weak O{\sc\,I} is given by a near-axis view of the bipolar/jet-like ionization structure. As the ionization decreases progressively with an increasing viewing angle with respect to the symmetry axis of the excitation front, the O{\sc\,I} opacity due to recombination will increase. Such a near-edge view would lead to a stronger and highly polarized O{\sc\,I}\,$\lambda$7774 as seen from SN\,2002ap~\citep{2003ApJ...592..457W}. 
Such a near-axis view of the excitation structure, which has been implemented in our modeling of the 3D Ca{\sc\,II} (i.e., a viewing angle of $\sim40^{\circ}$; see Section~\ref{sec:model}), also produces plausible fits to the observed flux and polarization spectra as illustrated in Figure~\ref{fig:model}. 
We also remark that an exact on-axis view will lead to a face-on viewing of the Ca{\sc\,II} opacity distribution, thus reducing the observed polarization due to a more symmetric configuration (see Appendix~\ref{sec:angle} and Appendix Figure~\ref{fig:angle}).

Alternatively, as discussed in the case study of SN\,2019ewu~\citep{2023ApJ...944L..49W}, a relatively low oxygen abundance of the progenitor that may be induced by a more effective stripping of the oxygen shell can naturally explain the observed weak O{\sc\,I}\,$\lambda$7774 line. Moreover, according to the simulations of~\citet{2012MNRAS.424.2139D}, a low degree of $^{56}$Ni in the ejecta will also produce a weak O{\sc\,I}\,$\lambda$7774 line. 
We note that SN\,2026gzf provides an example that is an exception to the general trend of stronger oxygen features in Type Ic than in Type Ib SNe~\citep{2001AJ....121.1648M, 2016ApJ...827...90L}. Detailed modeling of the spectrophotometric evolution of the SN is beyond the scope of this work. 

A bipolar explosion configuration was inferred from the anticorrelation between the line-profile shapes of the [Ca{\sc\,II}]\,$\lambda \lambda$7291,\,7323 and [O{\sc\,I}]\,$\lambda \lambda$6300,\,6363 emission features in the nebular spectra of stripped-envelope SNe~\citep{2024NatAs...8..111F}. The dual-component aspherical and axisymmetric excitation structure identified in SN\,2026gzf is compatible with this scenario. For an axisymmetric excitation structure, a flat-topped or horn-like profile of the [Ca{\sc\,II}] $\lambda\lambda$7291,7323 forbidden line can be expected in a spectrum of SN\,2026gzf at the nebular phase. 

Finally, we remark that the day 16.5 spectropolarimetry presented here does not directly constrain the properties of the shock breakout. However, the significantly asymmetric explosion, as traced by the excitation structure in the SN ejecta, strongly suggests that asymmetric shock breakout should be considered. 
Our 3D Monte-Carlo simulation finds a plausible fit to the observations by implementing an opacity enhancement confined within ranges of opening angle,  
which is generally compatible with an explosion induced by a jet, e.g., SNe\,2008D~\citep{2009ApJ...705.1139M} and 2014ad~\citep{2017MNRAS.469.1897S}.
A secondary component was introduced to interpret the departure from axisymmetry at $\sim28,000$\,km\,s$^{-1}$, which indicates a rather complex excitation geometry toward the outermost regions. 
Our constraint of a near-axis viewing angle of $\sim 40^{\circ}$ is consistent with the radio nondetections of SN\,2026gzf, which exclude an on-axis jet and restrict the viewing angle to $> 14^{\circ}\text{--}45^{\circ}$ depending on the assumed jet kinetic energy and environment \citep{2026arXiv260610014Y,2026arXiv260610002M}.
A recent VLT spectropolarimetry time sequence of the H-rich core-collapse SN\,2024ggi clearly revealed an axisymmetric shock breakout, the geometry of which is also shared by the expanding envelope of the SN~\citep{2025SciA...11x2925Y}. We therefore remark that semi-analytical modeling of the prompt shock-breakout emission under the framework of spherical symmetry should be considered with caution.

\begin{acknowledgments}
We thank the European Organisation for Astronomical Research in the Southern Hemisphere (ESO) for generous allocations of observing time. 
The imaging polarimetry and spectropolarimetry presented here is based on observations collected at ESO’s La Silla Paranal Observatory under programs ID 116.28T2.002 [PI M. Pursiainen] and 116.2ASW.001 [PI Y. Yang], respectively. 
We especially thank the staff of the Paranal Observatory for their proficient and diligent support of this project in service mode. 
Y.Y.'s research is partially supported by the Tsinghua University Dushi Program. 
J.L. acknowledges financial support from NSF-2206523. 
Q.J.H, B.W and X.f.W acknowledge support from the National Natural Science Foundation of China (grant No. 12321003). 
W.E.K. acknowledges financial support from NSF-2206523, NSF-2311323, and  HST-AR-16613. This research made use of \textsc{tardis}, a community-developed software package for spectral
synthesis in SNe \citep{2014MNRAS.440..387K, kerzendorf_2025_15069852}.
The development of \textsc{tardis} received support from GitHub, the Google Summer of Code
initiative, and ESA's Summer of Code in Space program. \textsc{tardis} is a fiscally
sponsored project of NumFOCUS. \textsc{tardis} makes extensive use of Astropy.
M.P. acknowledges support from a UK Research and Innovation Fellowship (UKRI1062). 
A.V.F. is grateful for financial assistance from numerous donors. 
G.L was supported by a research grant (VIL60862) from VILLUM FONDEN. 
S.S. is partially supported by LBNL Subcontract 7707915.

\end{acknowledgments}

\section*{Data Availability}
The data underlying this article will be shared on reasonable request to the corresponding author.

\begin{contribution}

X.~Wen developed the 3D Monte-Carlo Polarization Simulation Code, carried out the modeling of the spectropolarimetry data, contributed to the data interpretation, and helped develop the manuscript. 
Y.Y. is the PI of the VLT spectropolarimetry proposal, initiated the project, carried out the observations, reduced the data, conducted spectropolarimetric analysis, contributed to the data interpretation, and wrote the manuscript. 
J.L. conducted the {\sc TARDIS} modeling of the flux spectrum and helped develop the manuscript. 
L.W. and J.C.W. contributed to the data
interpretation and to the manuscript. 
\textcolor{black}{
M.P. is the PI of the VLT imaging polarimetry proposal, carried out imaging polarimetry observations, reduced the data, contributed to the design and execution of observations, and assisted with data interpretation.}
Q.H. identified the target and helped develop the VLT proposal. 
B.W. helped develop the VLT proposal. 
X.~Wu leads the high-energy time-domain astrophysics (HETIDA) group at Purple Mountain Observatory, helped develop the VLT proposal, and edited the manuscript.
A.V.F. leads the U.C. Berkeley supernova research group, contributed to the discussion, and edited the manuscript. 
W.E.K. leads the supernova research group at Michigan State University and developed the {\sc TARDIS} code for radiative transfer. 
G.L. contributed to the design and execution of the observations. 
S.S. and F.P. helped develop the VLT proposal and edited the manuscript.

\end{contribution}

\facility{VLT:Antu (FORS2)}

\software{astropy~\citep{2013A&A...558A..33A,2018AJ....156..123A,2022ApJ...935..167A},  
          PyRAF~\citep{2012ascl.soft07011S}, 
          TARDIS~(\citealt{2014MNRAS.440..387K}; version: \citealt{kerzendorf_2025_15069852})
          }

\setcounter{figure}{0}
\setcounter{table}{0}
\appendix

\section{Appendix}~\label{sec:Appendix information}

\subsection{TARDIS Modeling of the Day 16.5 Spectrum of SN~2026gzf}~\label{sec:tardis}
We used the open-source one-dimensional (1D) Monte-Carlo radiative-transfer code {\sc TARDIS}~(\citealt{2014MNRAS.440..387K}; version 2025.03.23\footnote{\url{https://github.com/tardis-sn/tardis/tree/release-2025.03.23}}: \citealt{kerzendorf_2025_15069852}) to model the flux spectrum of SN\,2026gzf on day 16.5 to constrain basic ejecta environment parameters. The modeling was carried out under the framework of a Bayesian inference that emulates {\sc TARDIS} radiative-transfer calculations with a probabilistic deep-learning model \citep{kerzendorf_probabilistic_2022}, following the prescription detailed by~\citet{2026ApJ..1002L..11L}. The emulator from~\citet{2026ApJ..1002L..11L} is trained on $700,000$ {\sc TARDIS} spectra assuming uniform fractional element abundances (except for Ca) and a power-law density profile ($\rho \propto r^{-n}$) in a high-dimensional parameter space optimized for SN\,2014L, a spectroscopically normal Type Ic SN. Here we adopt the same emulator for model-spectrum evaluation in the inference process, but allow the outer-edge ejecta velocity to vary over the prior range $25,000$ to $80,000$\,km\,$s^{-1}$, rather than fixing it at $35,000$\,km\,s$^{-1}$ as in~\citet{2026ApJ..1002L..11L}. Key {\sc TARDIS} parameter settings are listed in Table~\ref{Table_tardis_settings}.

\begin{figure*}[ht]
    \centering
    \includegraphics[trim={0.0cm 0.0cm 0.0cm 0.0cm},clip,width=1.0\textwidth]{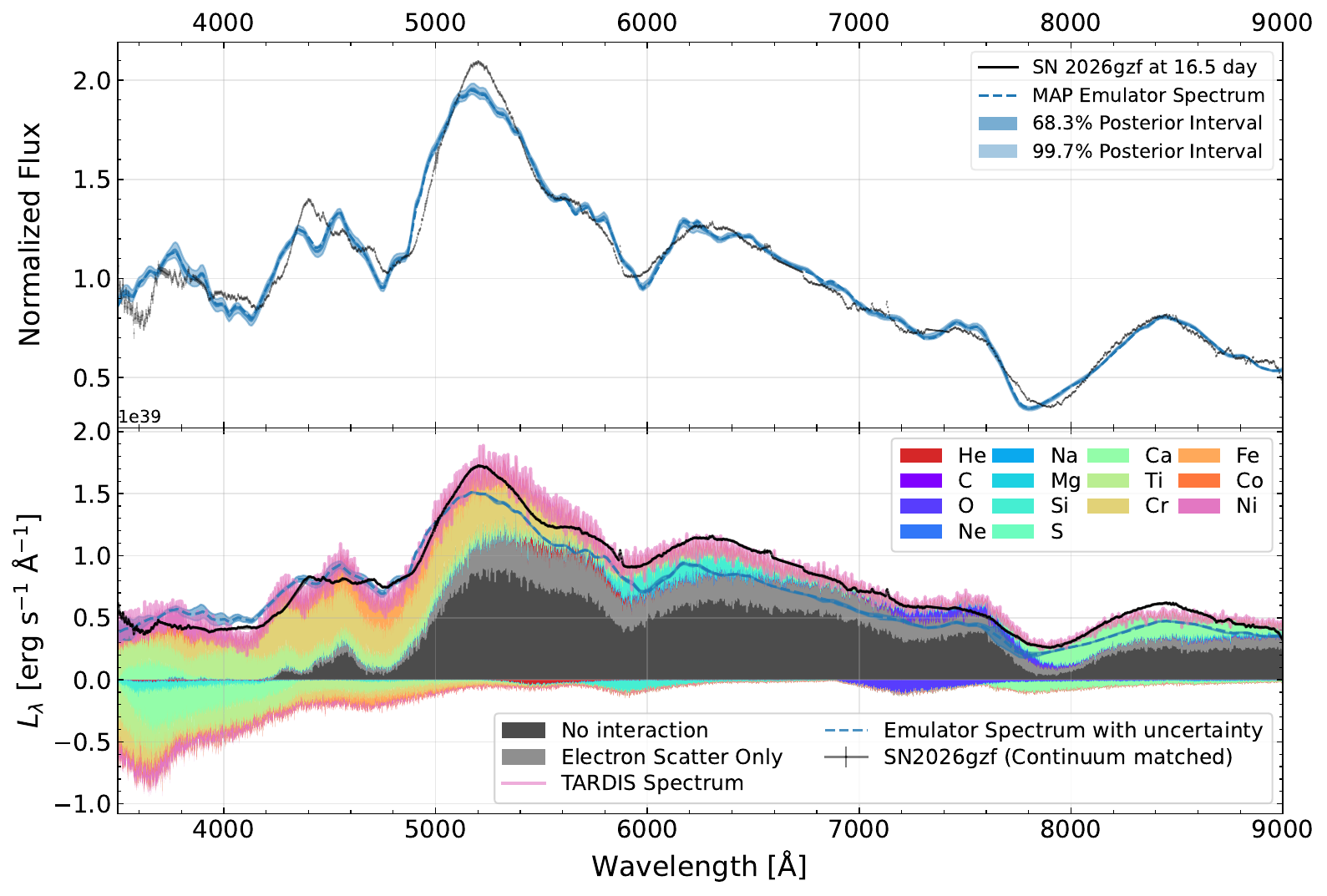}
    \vspace{-0.6 cm}
         \captionsetup{name=\suppmattwo Figure }
    \caption{
    The maximum-a-posteriori (MAP) emulator spectrum (top, navy curve) compared to the day 16.5 VLT spectrum of SN\,2026gzf (top, black curve). Narrow emission lines from underlying nebulosity were manually interpolated for the purpose of spectral modeling. The bottom panel compares the emulator spectra (navy dashed line) and the {\sc TARDIS} simulation (pink solid line) evaluated with the MAP parameter set. The continuum-matched observed spectrum of SN\,2026gzf is presented in black for comparison. The color-shaded spectra illustrate the elemental decomposition of the {\sc TARDIS} simulation. Black- and gray-shaded areas show contributions from noninteracting photons and those interacting only through electron scattering, respectively. 
    }
~\label{fig:tardis}
\end{figure*}

In Appendix Figure~\ref{fig:tardis}, we present our {\sc TARDIS} fit to the spectrum of SN\,2026gzf on day 16.5. The model reproduces the profiles of the primary absorption features, namely the Mg{\sc\,II}\,$\lambda$4481 and Si{\sc\,II}\,$\lambda$5051 respectively blended with several iron lines such as Fe{\sc\,II}\,$\lambda$4397 and Fe{\sc\,II}\,$\lambda$5018, $\lambda$5129. 
Other prominent silicon lines in the data, such as Si{\sc\,II}\,$\lambda$5972 and Si{\sc\,II}\,$\lambda$6355, are also well-reproduced in the model spectrum. We infer a power-law density exponent of $-6.41$ to $-6.14$ (16\% to 94\% credible interval) for the outer ejecta of SN\,2026gzf on day 16.5,  consistent with {\sc TARDIS} models of other Type Ic-BL SNe (see \citealp{2019Natur.565..324I, 2022ApJ...937...40K}). We note that the day 16.5 spectrum of SN\,2026gzf indicates a rather fast expanding ejecta, with a photospheric velocity $v_{\rm inner} = 18,807^{+110}_{-120}$ \,km\,s$^{-1}$, and an outer edge velocity of $v_{\rm outer} = 36,964 \pm 216$ \,km\,s$^{-1}$. The photospheric temperature is inferred to be $T_{\rm inner} = 8972 \pm 50 $\,K. The details of the spectroscopic time series modeling of SN\,2026gzf  phases will be discussed in a forthcoming paper.

\subsection{Continum Polarization Constraints on ejecta Geometry}~\label{sec:Continum Polarization}

\begin{figure*}[ht]  
    \centering
    \includegraphics[width=1.0\textwidth]{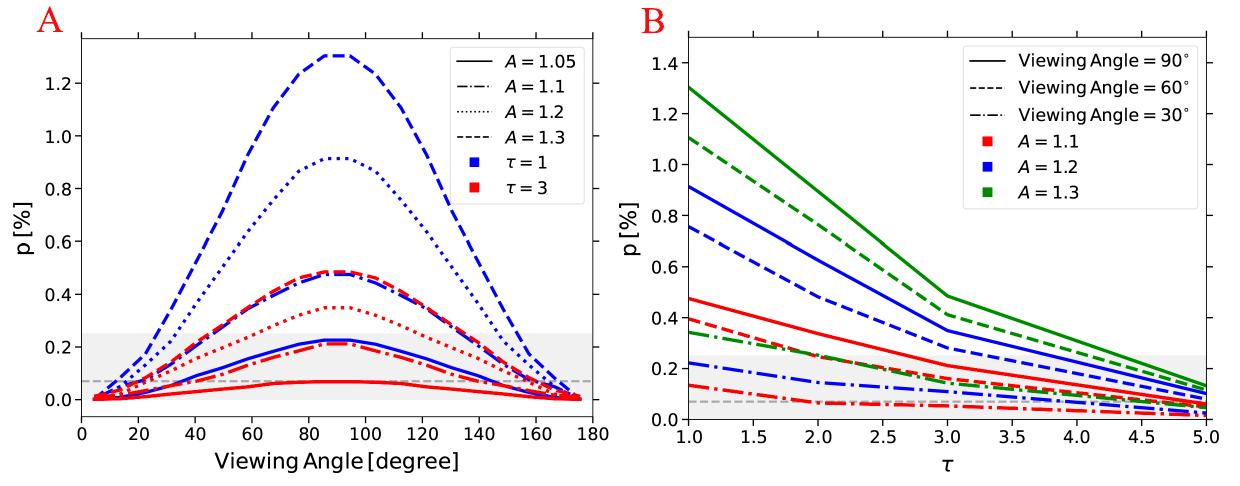}
    \vspace{-8.0mm}
     \captionsetup{name=\suppmattwo Figure }
    \caption{
Predicted continuum polarization from MCPOL simulations for axisymmetric ellipsoidal ejecta with different axial ratios and electron-scattering optical depths. {\it Panel A} shows the continuum polarization as a function of viewing angle for axial ratios $A=1.05, 1.1, 1.2, \, \rm{and}\, 1.3$. Blue and red curves correspond to photons emitted from electron-scattering optical depths of $\tau=1$ and $\tau=3$, respectively. The horizontal dashed line marks the observed bias-corrected continuum polarization, while the shaded region indicates its uncertainty. {\it Panel B} shows the continuum polarization as a function of electron-scattering optical depth for viewing angles of $30^{\circ}$, $60^{\circ}$ and $90^{\circ}$. Colors denote different axial ratios, while line styles represent different viewing angles.
}~\label{fig:con_pol} 
\end{figure*}

We used the MCPOL code to perform a suite of continuum polarization simulations across a range of parameter sets to constrain the allowed parameter space, adopting a method widely used in supernova polarization studies~\citep{1991A&A...246..481H,Kasen2003, 2020ApJ...902...46Y, 2023MNRAS.520..560H}. We adopted a prolate ellipsoidal power-law density distribution with a power-law exponent of $n = 6$. The degree of non-spherical symmetry is parameterized by the axial ratio $A = c/a$, ranging from approximately spherical ($A = 1.05$) to moderately distorted ($A = 1.3$). 
Because the polarization carried by escaping photons is determined primarily by their last few electron-scattering events, we follow \cite{Hoflich1993,Kasen2003} and inject initially unpolarized photon packets at an electron-scattering optical depth of $\tau \sim 3$, where the predicted polarization has approached its optically thick asymptotic value (as show in Figure~\ref{fig:con_pol}B). This value specifies the effective optical depth for photon injection and should not be interpreted as either the total center-to-edge optical depth of the ejecta.
For comparison, we also consider the case where photons are emitted at $\tau = 1$. The low polarization at day 16.5 favors a prolate configuration with $A \leq 1.05$, but the inferred asphericity is degenerate with both viewing angle and optical depth (as shown in Appendix Figure~\ref{fig:con_pol}A). 

As the ejecta expand and their density decreases, the region in which escaping photons undergo their last few electron scatterings recedes to lower velocities and deeper layers in mass coordinate. Time-dependent polarimetry can therefore probe the geometry of progressively deeper ejecta layers. 
The absence of a significant increase in continuum polarization between the two epochs disfavors highly aspherical configurations ($A > 1.2$), particularly for near edge-on viewing angles ($\theta \sim 90^\circ$). We therefore constrain the ejecta to exhibit only slight departures from spherical symmetry, with the degree of asymmetry strongly dependent on the viewing angle.

\subsection{Viewing-Angle Dependence of the Spectral and Polarization Profiles}~\label{sec:angle}

To systematically investigate how the Ca{\sc\,II}\,NIR3 line profiles and percent polarization  depend on the viewing angle, we construct a simplified reference model featuring a single bipolar enhancement. This control component adopts identical physical parameters with the primary component described in Section~\ref{sec:model}, except that its symmetry axis is aligned with the $z$ axis (as illustrated in Appendix Figure~\ref{fig:angle}A). 
Such an adjustment of the symmetry axis would project the model polarization of our specified axisymmetric configuration onto the $\rm Q$ axis. As the polarization on $\rm U$ vanishes, the model polarization spectrum of $\rm Q$ can be directly compared to the observed polarization spectrum ($\rm p$).

As the viewing angle shifts from the equator to the pole, the simulated spectral profile of Ca{\sc\,II}\,NIR3 exhibits deeper absorption at the blue end and a rise at the red end. Concurrently, the percent polarization 
varies nonmonotonically, characterized by an initial increase followed by a subsequent decline, with its peak migrating toward higher velocities. This behavior arises because, as the LOS approaches $0^\circ$, high-velocity Ca{\sc\,II} material increasingly occults the photosphere, boosting the net polarization and shifting the peak toward higher velocities. Near the polar region, however, the enhanced geometric symmetry along the LOS leads to significant polarization cancellation, causing the percent polarization 
to decrease. 
Appendix Figure~\ref{fig:angle}B illustrates that for a $\lesssim30^{\circ}$ viewing angle, the model-synthesized Ca{\sc\,II}\,NIR3 line profile shows apparent deviation from the data toward the red end.
A $\gtrsim60^{\circ}$ viewing angle will produce significantly lower line polarization compared to the observation, as presented in Appendix Figure~\ref{fig:angle}C.
A simultaneous fit to both the spectral and polarization profiles can be achieved at a moderately off-axis viewing angle of $\sim 40^\circ$.

\begin{figure*}[ht]  
    \centering
    \includegraphics[width=1.0\textwidth]{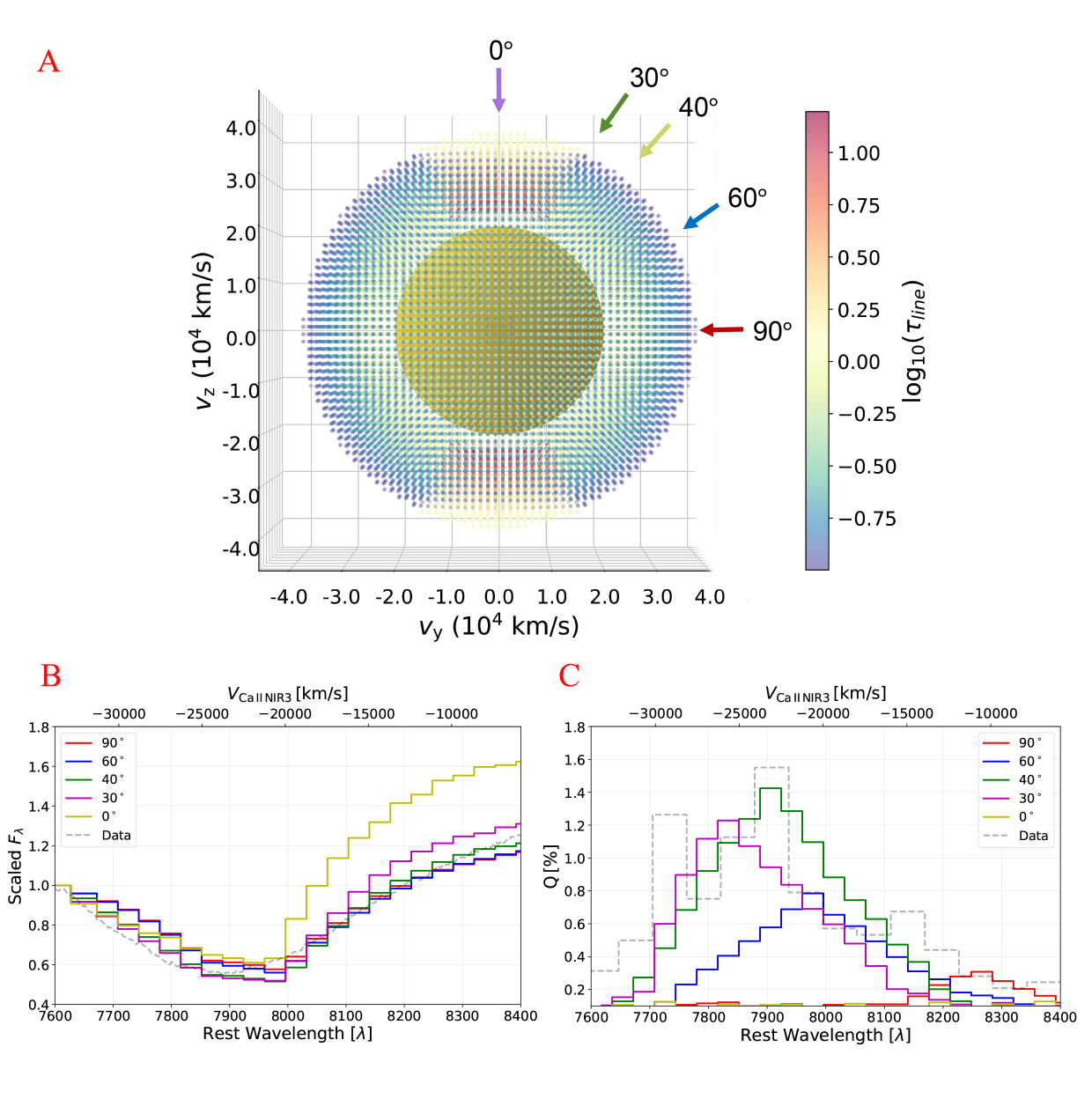}
    \vspace{-8.0mm}
     \captionsetup{name=\suppmattwo Figure }
    \caption{
Investigating the viewing-angle dependence from the flux and polarization profiles across the Ca{\sc\,II}\,NIR3 feature. 
{\it Panel A} illustrates the opacity enhancement within a cone-shaped region as described in the text. {\it Panel B} compares the observed spectral profile with a series of model-synthesized flux density ($F_{\lambda}$) as observed at different inclination angles as indicated by the legend. The corresponding polarization spectra projected onto the $\rm Q$ axis are presented in {\it Panel C}. Velocities measured with respect to the center wavelength of Ca{\sc\,II}\,NIR3 can be read off the top abscissas. Small fluctuations in the histograms that show the simulation results are due to Monte-Carlo statistical noise.
}~\label{fig:angle} 
\end{figure*}

\setlength{\tabcolsep}{3.5pt}
\begin{table}
\begin{center}
     \captionsetup{name=\suppmattwo Table }
\caption{SN\,2026gzf {\it TARDIS} model settings.~\label{Table_tardis_settings}}
\begin{normalsize}
\begin{tabular}{c|c}
\hline
\hline
{\it TARDIS} setting  &  Values   \\
\hline
Atomic Data      &  kurucz\_cd23\_chianti\_H\_He.h5 \\
Ionization       &  nebular \\
Excitation       &  dilute-lte \\
Radiative Rate   &  dilute-blackbody \\
Line Interaction &  macroatom  \\
Helium Treatment &  recomb-nlte \\
\hline
\hline
\end{tabular}
\end{normalsize}
\end{center}
\end{table}


\bibliography{specpol}{}
\bibliographystyle{aasjournalv7}



\end{document}